\documentstyle[rotating]{mn}	

\newif\ifAMStwofonts

\def\refit{}\def\refbf{}
\def\myref#1,#2,#3.{{\refit #1\/}, {\refbf #2}, #3\par\noindent}
\def\aa,#1,#2.{\myref{A\&A},#1,#2.}
\def\acta,#1,#2.{\myref{Acta Astron.},#1,#2.}
\def\annrev,#1,#2.{\myref{ARAA},#1,#2.}
\def\aj,#1,#2.{\myref{AJ},#1,#2.}
\def\apj,#1,#2.{\myref{ApJ},#1,#2.}
\def\apjsupp,#1,#2.{\myref{ApJS},#1,#2.}
\def\apspsci,#1,#2.{\myref{Ap\&SS},#1,#2.}
\def\aasupp,#1,#2.{\myref{AA\ Supp.},#1,#2.}
\def\ica,#1,#2.{\myref{Icarus},#1,#2.}
\def\grg,#1,#2.{\myref{GRG},#1,#2.}
\def\jaa,#1,#2.{\myref{J.\ Astr.\ Astrophys.},#1,#2.}
\def\mnras,#1,#2.{\myref{MNRAS},#1,#2.}
\def\nat,#1,#2.{\myref{Nature},#1,#2.}
\def\pasp,#1,#2.{\myref{PASP},#1,#2.}
\def\pasj,#1,#2.{\myref{PASJ},#1,#2.}
\def\physrev,#1,#2.{\myref{Phys.\ Rev.},#1,#2.}
\def\physrevlett,#1,#2.{\myref{Phys.\ Rev.\ Lett.},#1,#2.}
\def\physrevD,#1,#2.{\myref{Phys.\ Rev.\ D},#1,#2.}
\def\procroy,#1,#2.{\myref{Proc.\ Roy.\ Soc.},#1,#2.}
\def\revmod,#1,#2.{\myref{Rev.\ Mod.\ Phys.},#1,#2.}
\def\sova,#1,#2.{\myref{SvA},#1,#2.}

\def\etal{{\it et\thinspace al\/}}

\def\ie{{i.e}}
\def\eg{{e.g}}

\def\cf{{cf}}
\def\kms{\mbox{km s$^{-1}$}}
\def\mpc{\mbox{Mpc}}

\def\frac(#1/#2){{\textstyle{#1\over#2}}}

\def\subrm#1{_{\rm #1}}
\catcode`|=\active \let|=\subrm

\def\pd#1#2{{\upartial#1\over \upartial#2}} \def\lr#1{\left(#1\right)}

\newcount\psw 
\def\computelabs{
  \psw=\number\epsfxsize \divide\psw by 65536 
}%
\newbox\labelbox
\newdimen\yy\newdimen\xx
\newdimen\hair\hair=3pt

\def\xlabel#1#2#3{%
\setbox\labelbox\hbox{$#1$}%
\getkerns{#2}{#3}%
\advance\xx by -.5\wd\labelbox%
\putlabel{0}%
}

\def\ylabel#1#2#3{%
\setbox\labelbox\hbox{$#1$}%
\getkerns{#2}{#3}%
\advance\yy by .5\wd\labelbox%
\putlabel{90}%
}

\def\putlabel#1{%
\vbox to 0pt{\kern-\yy\hbox to 0pt{\kern\xx%
\rotatebox{#1}{\box\labelbox}%
\hss}\vss}%
\ifvmode\nointerlineskip\fi%
}

\def\getkerns#1#2{%
\yy=#2 pt 
\divide \yy by \mag%
\multiply \yy by 1010%
\multiply \yy by \the\psw%
\xx=#1 pt 
\divide \xx by \mag%
\multiply \xx by 1010%
\multiply \xx by \psw%
\advance \xx by .5\hsize
\advance \xx by -.5\epsfxsize
}

\def\reference{\item}
\def\vmax{v|{m}}
\def\rd{R|d}
\def\rh{R|h}
\def\MI{{\cal M}_I}
\def\epsl{\epsilon_{l}}
\def\epsd{\epsilon|{d}}

\def\epsm{\epsilon|{m}}
\def\lm{\lambda|{m}}
\def\am{a|{m}}
\def\m2l{\Upsilon_I}
\def\bv{B\hbox{$-$}V}

\def\Msun{{\ifmmode M_{\sun} \else $M_{\sun}$ \fi}}
\def\msun{{\ifmmode m_{\sun} \else $m_{\sun}$ \fi}}
\def\Lsun{{\ifmmode L_{\sun} \else $L_{\sun}$ \fi}}
\def\rsun{{\ifmmode r_{\sun} \else $r_{\sun}$ \fi}}
\def\Rsun{{\ifmmode R_{\sun} \else $R_{\sun}$ \fi}}
\def\pc{{\ifmmode \hbox{pc} \else {pc} \fi}}

\def\MMWr{Mo, Mao \& White (1998)}
\def\MMW{MMW}

\input epsf

\ifoldfss
  \ifCUPmtlplainloaded \else
    \NewTextAlphabet{textbfit} {cmbxti10} {}
    \NewTextAlphabet{textbfss} {cmssbx10} {}
    \NewMathAlphabet{mathbfit} {cmbxti10} {} 
    \NewMathAlphabet{mathbfss} {cmssbx10} {} 
  \fi
  \ifAMStwofonts
    \ifCUPmtlplainloaded \else
      \NewSymbolFont{upmath} {eurm10}
      \NewSymbolFont{AMSa} {msam10}
      \NewMathSymbol{\upi}     {0}{upmath}{19}
      \NewMathSymbol{\umu}     {0}{upmath}{16}
      \NewMathSymbol{\upartial}{0}{upmath}{40}
      \NewMathSymbol{\leqslant}{3}{AMSa}{36}
      \NewMathSymbol{\geqslant}{3}{AMSa}{3E}

    \fi
  \fi
\fi 

\ifnfssone
  \newmathalphabet{\mathit}
  \addtoversion{normal}{\mathit}{cmr}{m}{it}
  \addtoversion{bold}{\mathit}{cmr}{bx}{it}
  \newmathalphabet{\mathbfit} 
  \addtoversion{normal}{\mathbfit}{cmr}{bx}{it}
  \addtoversion{bold}{\mathbfit}{cmr}{bx}{it}
  \newmathalphabet{\mathbfss} 
  \addtoversion{normal}{\mathbfss}{cmss}{bx}{n}
  \addtoversion{bold}{\mathbfss}{cmss}{bx}{n}
  \ifAMStwofonts
    \ifCUPmtlplainloaded \else
      \UseAMStwoboldmath
      \makeatletter
      \new@mathgroup\upmath@group
      \define@mathgroup\mv@normal\upmath@group{eur}{m}{n}
      \define@mathgroup\mv@bold\upmath@group{eur}{b}{n}
      \edef\UPM{\hexnumber\upmath@group}
      \new@mathgroup\amsa@group
      \define@mathgroup\mv@normal\amsa@group{msa}{m}{n}
      \define@mathgroup\mv@bold\amsa@group{msa}{m}{n}
      \edef\AMSa{\hexnumber\amsa@group}
      \makeatother
      \mathchardef\upi="0\UPM19
      \mathchardef\umu="0\UPM16
      \mathchardef\upartial="0\UPM40
      \mathchardef\leqslant="3\AMSa36
      \mathchardef\geqslant="3\AMSa3E
    \fi
  \fi
\fi 

\ifnfsstwo
  \DeclareMathAlphabet{\mathbfit}{OT1}{cmr}{bx}{it}
  \SetMathAlphabet\mathbfit{bold}{OT1}{cmr}{bx}{it}
  \DeclareMathAlphabet{\mathbfss}{OT1}{cmss}{bx}{n}
  \SetMathAlphabet\mathbfss{bold}{OT1}{cmss}{bx}{n}
  \ifAMStwofonts
    \ifCUPmtlplainloaded \else
      \DeclareSymbolFont{UPM}{U}{eur}{m}{n}
      \SetSymbolFont{UPM}{bold}{U}{eur}{b}{n}
      \DeclareSymbolFont{AMSa}{U}{msa}{m}{n}
      \DeclareMathSymbol{\upi}{0}{UPM}{"19}
      \DeclareMathSymbol{\umu}{0}{UPM}{"16}
      \DeclareMathSymbol{\upartial}{0}{UPM}{"40}
      \DeclareMathSymbol{\leqslant}{3}{AMSa}{"36}
      \DeclareMathSymbol{\geqslant}{3}{AMSa}{"3E}
    \fi
  \fi
\fi 

\ifCUPmtlplainloaded \else
  \ifAMStwofonts \else 
    \def\upi{\pi}
    \def\umu{\mu}
    \def\upartial{\partial}
  \fi
\fi

\title[] 
{Observational constraints on disk galaxy formation}
\author[]
{D. Syer, Shude Mao, and H.J. Mo
\thanks{E-mail: (syer, smao, hom)@mpa-garching.mpg.de} \\
	Max-Planck-Institut f\"ur Astrophysik
	Karl-Schwarzschild-Strasse 1, 85748 Garching, Germany}
\date{Accepted ........
      Received .......;
      in original form .......}

\pagerange{\pageref{firstpage}--\pageref{lastpage}}
\pubyear{1997}

\begin{document}
\maketitle
\label{firstpage}

\begin{abstract}
We use data from the literature to constrain theoretical models of
galaxy formation.  We show how to calculate the dimensionless spin
parameter $\lambda$ of the halos of disk galaxies and we compare the
distribution of $\lambda$ with that observed in cosmological $N$-body
simulations.  The agreement is excellent, which provides strong
support for the hierarchical picture of galaxy formation.  Assuming
only that the radial surface density distribution of disks is
exponential, we estimate crudely the maximum-disk mass-to-light ratio
in the $I$-band and obtain $\langle\m2l\rangle \la 3.56h$, for a Hubble
constant of $100h~\kms \mpc^{-1}$.  We discuss this result and its
limitations in relation to other independent determinations of $\m2l$.
We also define a dimensionless form of the Tully-Fisher relation, and
use it to derive a value of the baryon fraction in disk galaxies.  For
galaxies with circular velocity $\vmax>100\kms$, the median value is
$m|d = 0.086 (\,{\m2l/3.56h})$.  Assuming that the gas fraction in
galactic halos is at most as large as that in clusters, we also
conclude that $\langle\m2l\rangle \la 2.48h^{-1/2}$.
\end{abstract}


\section{Introduction}
The fact that the rotation curves of spiral galaxies are rather flat
is usually taken to imply the presence of an extended halo of dark
matter (\eg{.} Freeman 1970, Persic \& Salucci 1991).  If such dark
matter exists, then hierarchical models of galaxy formation (White \&
Rees 1978) are a natural consequence of gravitational instability.  In
these models the standard picture of disk formation is that gas, which
is initially distributed in the same way as the dark matter, cools and
settles into rotationally supported disks at the centres of 
dark matter halos (Fall \& Estafthiou 1980).

Dalcanton, Spergel \& Summers (1997), and \MMWr{} (\MMW) have shown
recently that such a picture can reproduce some of the broad
properties of observed disk galaxies.  In particular they show that
the Tully-Fisher relation (Tully \& Fisher 1977) can be understood
very simply.  However, as was pointed out by Courteau \& Rix (1997),
the Tully-Fisher relation is incompatible with the
maximum-disk hypothesis (Carignan \&
Freeman 1985) if disk galaxies have universal mass-to-light ratios
(de Jong 1996). 

Great interest in the Tully-Fisher relation as a distance indicator
(Giovanelli \etal{.}  1997 and references therein) has led to many
observations of disk galaxies, and large samples are now available
(\eg{.} Mathewson \& Ford 1996; Courteau 1996, 1997). In this paper,
we examine the observational 
constraints which can be placed on the standard picture
of disk formation.

In the next section we review the properties of exponential disks
and the scenario of disk formation. We start from the standard
assumption that the disk mass-to-light ratio
$\Upsilon$ is universal (or at least does not vary
strongly with surface brightness), and calculate
various components from observable quantities.  
In Section \ref{obsec} we describe the observations of disk galaxies.
In Section \ref{galsec} we show the results of applying the theory of Section
\ref{modelsec} to the data.  In Section \ref{dissec} we discuss the
implications of our results and draw conclusions.  In particular,
Section \ref{detsec} is devoted to a discussion of independent
determinations of $\Upsilon$.

\section {Disk Formation}\label{modelsec}

\subsection{Exponential disks}\label{expsec}

The luminous disks of spiral galaxies are commonly modelled by an
exponential surface brightness distribution:
\begin{equation}
\mu(R) = {L|d\over2\pi\rd^2} \exp(-R/\rd)
\label{muexp}
\end{equation}
where $R$ is the usual cylindrical radius, 
$R|d$ is the exponential scalelength, and $L|d$ is the total
luminosity of the disk.  Here we collect some notation and a number of
useful results relating to exponential disks.

The disk has a mass $M|d$, and a mass-to-light ratio 
$\Upsilon$ in solar units.  Thus the surface mass density of the disk is
\begin{equation}
\Sigma(R) = \mu(R)\Upsilon = {M|d\over2\pi\rd^2} \exp(-R/\rd).
\label{Sigexp}
\end{equation}
The gravitational potential in the disk $\Phi$
is conveniently decomposed into contributions from the disk and a halo:
\begin{equation}
\Phi = \Phi|{d} + \Phi|{h}.
\label{phid}
\end{equation}
(We use the subscripts `d' for `disk', and `h' for `halo' throughout.)
We assume for the present purposes that the halo is spherical, and
usually we think of it as being composed of dark matter, but it may
also contain a stellar component (\eg{.} the `bulge' of an earlier
type spiral).

The speed of test particles on circular orbits $v|c$ as a function of
$R$ is given by
\begin{equation}
v^2|c(R) = -R\pd{\Phi}{R}.
\label{rotc}
\end{equation}
We shall refer to $v|c(R)$ as the rotation curve of the system.  The
rotation curve as measured in HI (apart from small contributions from
turbulent motion) is thought to be a good measure of the true rotation
curve as long as the system is axisymmetric.  

We characterise the self gravity of the disc through the the
dimensionless quantity
\begin{equation}
\epsm = {\vmax\over(GM|d/\rd)^{1/2}},
\label{epsdef}
\end{equation}
where $\vmax$ is the maximum value of $v|c$, and $G$ is the
gravitational constant.  The rotation curve of an isolated exponential
disk ($\Phi|h=0$) is given by Freeman (1970).  An isolated disk has
$\epsm=\epsd\approx0.63$ and a disk embedded in a halo has
$\epsm>\epsd$.

The directly observable quantity corresponding to $\epsm$ is
\begin{equation} 
\epsl = {\vmax \over (GL|d/\rd)^{1/2}},
\label{epsl}
\end{equation}
which is related to $\epsm$ by 
\begin{equation}
\epsl=\epsm\Upsilon^{1/2}.
\label{epslm}
\end{equation}
The quantity $\epsl^2$ has the units of a mass-to-light ratio, and
indeed it is a measure of the {\em total} mass (including dark matter
halo) contributing to the rotation curve.  Let us define a quantity
\begin{equation} 
\Upsilon^{\rm tot}(R) = {v^2|{c}(R) R  \over GL(R)},
\label{upstot}
\end{equation}
which measures the total mass-to-light ratio as a function of radius.
For an isolated disk it is a constant ($=\Upsilon$) and with an
extended dark halo it increases with radius.  The maximum rotation
velocity in general occurs at $R=R|{max}>\rd$, and the luminosity
enclosed is $L|{max}<L|d$, hence $\epsl^2<\Upsilon^{\rm tot}(R|{max})$.
For an isolated disk $R|{max}=2.2\rd$ and $L|{max}=0.65L|d$, so
$\Upsilon=\Upsilon^{\rm tot}(R|{max})=3.4\epsl^2$.

\subsection{Disk Formation Model}\label{mmwsec}

Here we reproduce a simple model where the dark matter halo is assumed
to be a singular isothermal sphere and disc self-gravity is neglected
(\cf{} MMW). 
In this model, the disk scale
length is
\begin{equation}
\rd = {1\over \sqrt{2}}\lambda \rh
\label{rdmmw}
\end{equation}
where $\lambda$ is the dimensionless spin parameter, $\rh$ is the
virial radius of the halo. The halo properties are given by
\begin{equation}
M|h = {v|h^2 \rh\over G},~~ \rh = \chi\, {v|h\over H_0 }
\label{rhmhmmw}
\end{equation}
where $M|h$ is the mass and $v|h$ the circular velocity of the halo,
$H_0=100h \kms \mpc^{-1}$ is the Hubble constant and $\chi$ is a
dimensionless constant. For discs assembled at redshift $z$, MMW argue
that $\chi\approx 0.1 H_0/H(z)$, where $H(z)$ is the Hubble constant
at $z$. We can treat $\chi$ as an adjustable parameter and derive its
value from observational data.  Whenever we need a numerical value for
$\chi$ we use the one derived in Section \ref{spinsec} ($\chi=0.049$).
The maximum rotation velocity of the disk is $\vmax \approx v|h$.

When more realistic halo profiles are used and disc self-gravity is
taken into account, the relation between $R|d$ and $v|h$ is slightly
modified from that given by equations (\ref{rdmmw}) and
(\ref{rhmhmmw}) (see MMW for details).  The constant part of such
modifications can be taken into account by the constant $\chi$. For
simplicity, we will ignore all high order effects.

The disk central surface density is given by
\begin{equation}
\Sigma_0 ={ m|d M|h\over2\pi\rd^2}
\label{mdmmw}
\end{equation}
where $m_d$ is the fraction of halo mass that settles into the disk.
We now define a dimensionless ratio of observables
\begin{eqnarray}
\lm^2 & \equiv & {H_0\over h} {\vmax\over\pi G \mu_0} 
{1\over\Upsilon}\nonumber\\
& = & {1.47\times10^{-2}}\; 
\lr{v|m\over200\kms}\; \lr{100\Lsun\pc^{-2} \over \mu_0}
\label{lamdef}
\end{eqnarray}
where $\mu_0=\Sigma_0/\Upsilon$ is the central surface brightness.
Defining also the structural parameter
\begin{equation}
\am \equiv {m|d h\over \Upsilon}\;{1\over\chi}\;,
\label{amdef}
\end{equation}
we combine equations (\ref{rdmmw}-\ref{amdef}) to obtain an expression
for $\lambda$:
\begin{equation}
\lambda^2 = \lm^2 \;\am.
\label{lam}
\end{equation}

{}From equations (\ref{epsdef}), (\ref{rdmmw}) and
(\ref{rhmhmmw}) we find that
\begin{equation} 
\epsm^2 = {1\over \sqrt{2}}{\lambda\over m|d}
\label{tfdim}
\end{equation}
which is a dimensionless description of disk formation: the smaller
the spin parameter $\lambda$, the more concentrated the disk, and the
more self-gravitating it is (smaller $\epsm$).  We can write equation
(\ref{tfdim}) in terms of observables as
\begin{equation} 
\epsl^2 h^{-1} =  {1\over \sqrt{2}\chi} \; { \lm \over\am^{1/2}}.
\label{epslam}
\end{equation}
Equation (\ref{epslam}) is just the dimensionless form of the
Tully-Fisher relation.  To see this we write the Tully-Fisher relation
as
\begin{equation} 
H_0^2 L = A v|m^3,
\label{tf}
\end{equation}
where the value of $A$ can be derived from equations
(\ref{rdmmw}-\ref{mdmmw}):
\begin{equation} 
A = \am \; \chi^2\;{H_0\over h G}.
\label{Adef}
\end{equation}
Equations (\ref{epslam}) and (\ref{tf}) both relate observable
quantities via the same constant of proportionality
(i.e. $\am\chi^2$), and hence
(\ref{epslam}) is a dimensionless form of the Tully-Fisher relation
(\ref{tf}). 

\begin{figure}
\epsfxsize=.9\hsize
\computelabs
\large
\ylabel{\lm}{-.0}{-.5}
\xlabel{\mu_0 (\hbox{mag/arcsec}^2)}{.5}{-.95}
\normalsize
\centerline{\epsfnormal{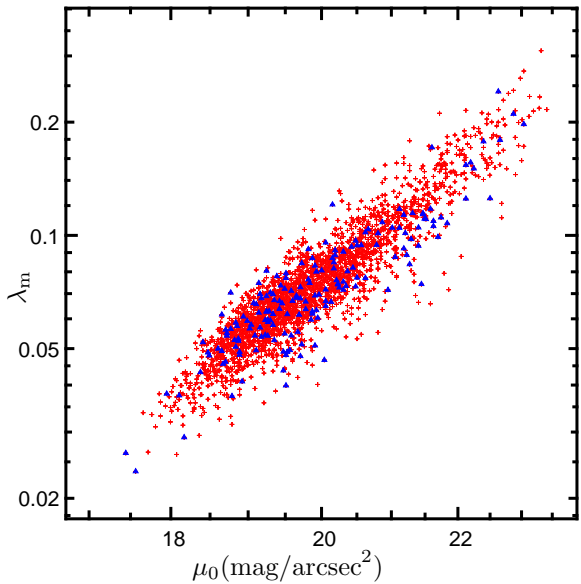}}
\vskip\baselineskip
\caption{ The relationship between $\lm$ and the central surface
brightness $\mu_0$ (in the I band)
of the sample of spiral galaxies of Mathewson \&
Ford (1996).  Barred galaxies are shown as triangles. }
\label{lamufig}
\end{figure}

\begin{figure}
\epsfxsize=.9\hsize
\computelabs
\large
\ylabel{\epsl^2/ h}{-.0}{-.5}
\xlabel{\lm}{.5}{-.95}
\normalsize
\centerline{\epsfnormal{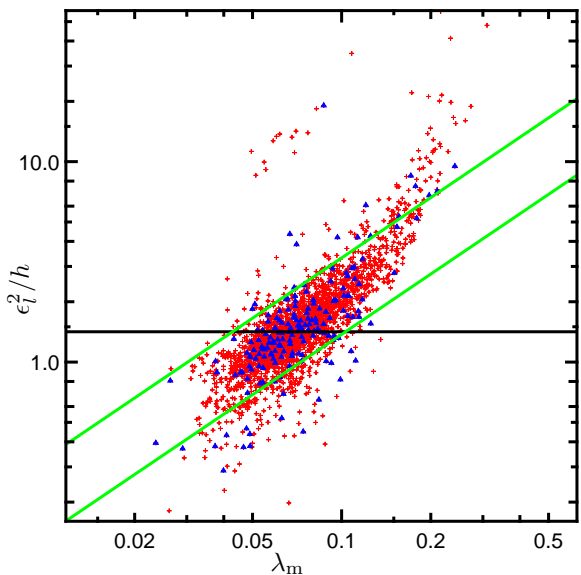}}
\vskip\baselineskip
\caption{ The dimensionless Tully-Fisher relation (equation
\ref{epslam}) for the MF data set.  On this plot, a fixed value of
$\epsm$ corresponds to a horizontal line with amplitude $\epsm^2(\m2l
h^{-1})$.  The horizontal solid line shows the value of $\epsm$ for a
self-gravitating disk ($\epsm=0.63$) with $\m2l=3.56h$. The two dashed
lines contain 90\% of the data with slope and scatter predicted by the
Tully-Fisher relation, within the framework of \MMW{} (see text).
Barred galaxies are shown as triangles. }
\label{datfig}
\end{figure}

An important parameter in the above model is the dimensionless spin
$\lambda$ of the system before disk formation.  In the hierarchical
model of galaxy formation, much is known about its expected value and
distribution (Cole \& Lacey 1996, Warren \etal{.} 1992,
Lemson \& Kauffmann 1998).  We can use equations (\ref{lam}) and
(\ref{epslam}) to write
\begin{equation}
\lambda = {1\over \sqrt{2}\chi} {\lm^2 h \over \epsl^2}
\label{lamchi}
\end{equation}
the right hand side of which contains {\em only} observable quantities
(and the constant $\chi$). Thus, this equation can be used to derive
the distribution of $\lambda$ from observations once the value of
$\chi$ is fixed (see Section \ref{spinsec}).

\section{Observations}\label{obsec}
To compare theory with observations we use the dataset of Mathewson \&
Ford (1996) (MF) which has rotation velocities and $I$-band photometry
for a sample of nearly 2500 Southern spiral galaxies selected randomly
from the ESO-Uppsala catalogue (Lauberts 1982).  The majority are
relatively late types: of those 2275 for which Hubble types are given,
1055 are Sc or Sbc; 814 are Sb; and 5 are Sa.  We convert the
published photometric quantities to $\rd$ and the central surface
brightness $\mu_0$ by assuming an exponential profile.  Details are
given in an Appendix.  

The observed Tully-Fisher relation (Giovanelli
et al. 1997; Shanks 1997):
\begin{equation} 
\MI - 5\log h = -(21.00 \pm 0.02) - (7.68\pm 0.13) (\log W - 2.5),
\label{TF}
\end{equation}
where $W$ is the inclination-corrected width of the HI line profile.
The MF data has a Tull-Fisher relation which is consistent with this,
albeit with larger scatter than in more carefully selected samples.

The maximum of the rotation curve $\vmax$ is given by $\vmax=W/2$.
Figure \ref{lamufig} shows the relationship between $\lm$ and the
central surface brightness $\mu_0$ of the MF data set.  The
correlation is expected, since halos with smaller $\lambda$ should
form more compact disks with higher $\Sigma_0$.  This figure
illustrates that $\lambda$ and $\mu_0$ are almost interchangeable,
despite the fact that in principle the correlation could have been
washed out by scatter in $\vmax$. The tight correlation between $\lm$
and $\mu_0$ therefore implies that the disk central surface brightness
is determined mainly by halo spin parameter rather than by halo
circular velocity, as is expected in the disk formation model (see \S
\ref{mmwsec}). {}From equations (\ref{rdmmw})-(\ref{mdmmw}) we see
that the disk central surface density scales with $v|h$ and $\lambda$
as $\Sigma_0\propto v|h/\lambda^2$.  The dynamical range of $v|h$ for
disk galaxies is a factor of about 3, while that for $\lambda^2$
predicted for dark halos is a factor of about 15.
 
In Figure \ref{datfig}, we show $\epsl$ as a function of $\lm$ for the
MF data. The extra factor of $h$ in the abscissa makes the plotted
quantities independent of the Hubble constant.  The figure reveals a
marked correlation between $\epsl$ and $\lm$ which corresponds to the
dimensionless Tully-Fisher relation (equation \ref{epslam}).  The slope
is unity provided $\am$ is independent of $\lm$.  The figure shows
that the slope is consistent with unity for the majority of galaxies.
At the large $\lm$ end the observed $\epsl$ is slightly
higher than the model prediction. There appears to be a break in the
slope of $\epsl$ versus $\lm$ at $\lm\approx0.1$.  
According to the model in Section
\ref{mmwsec} a higher value of $\epsl$ would mean that the combination
$m_d h/\Upsilon$ is lower for low-surface-brightness
galaxies.  This would result from a lower star formation efficiency
(hence higher $\Upsilon$), as suggested by observations (e.g. McGaugh
\& de Blok 1997).

For a given $\lm$, the scatter in $\epsl$ is determined by the scatter
in the Tully-Fisher amplitude $A$. In Figure \ref{datfig}, we overlay
the predicted slope (\ie{.} unity, see equation \ref{epslam}) and
scatter on top of the data points with the normalisation chosen to
reproduce the observed median value of $\epsl$ in the range
$\lm\in(.05,.06)$.  The lines are derived from the 5\% and 95\%
quartiles of $A$ of the data---the equivalent scatter in $A$ is a
factor of $\approx 2.4$ or $0.95$ magnitudes.  The predicted slope and
scatter are consistent with the observed data points. Thus, the data
are consistent with the assumption that the value of $\am$ is
independent of $\lm$ for the majority of these galaxies.  This is an
important result, because it means that the total mass-to-light ratio,
$M|h/L|d=\Upsilon/m|d\propto \am^{-1}$, is a constant for the majority
of galaxies. We will return to this point in the next section.

\section{Constraining Galaxy Formation}\label{galsec}
In this Section, we use the observational data to constrain
the disk model described in Section \ref{mmwsec}. 
We concentrate on the three most important parameters 
in the disk model: the mass-to-light ratios of disks,
the spin parameters of halos, and the baryon fractions
in disks. 

\begin{figure}
\epsfxsize=.9\hsize
\computelabs
\large
\ylabel{P(\m2l/h)}{-.02}{-.5}
\xlabel{\m2l/h}{.5}{-.95}
\normalsize
\centerline{\epsfnormal{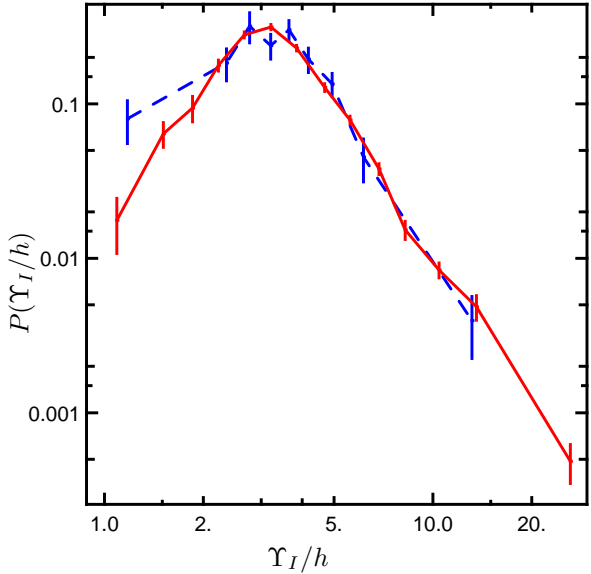}}
\vskip\baselineskip
\caption{
The distribution of $\m2l$ for the MF data, on the assumption that all
disks are exponential and self-gravitating.  The solid line is the
whole sample, and the dashed line is for the barred subsample.
The error bars are estimated using a standard bootstrap 
method.}
\label{ydist}
\end{figure}

\subsection{Mass-to-light ratios}\label{upssec}
Assuming that the disks are exponential sets a limit on $\m2l$: they
should all have $\epsm>0.63$, the value for an isolated disk.  Thus
for each galaxy $\m2l\la (\epsl/0.63)^2$, which is a crude
estimate of the maximum-disk mass-to-light ratio.  Figure \ref{ydist}
shows the distribution of $\m2l$ calculated in this way for the MF
data, and for the barred subset.  In Figure \ref{datfig} the solid
horizontal line marks the median of $\epsl$ in the data of Mathewson
\& Ford (1996).  This corresponds to $\epsm=0.63$ for
$\m2l=3.56h$.

A conservative upper limit for the average disk galaxy would be
$\langle\m2l\rangle<3.56h$.  This limit is also consistent with
independent measurements of $\m2l$ (see Section 5.1 for
details). However, since $\epsl$ is higher for low surface brightness
galaxies, a higher $\m2l$ is still allowed for these galaxies without
violating the constraint $\epsm>0.63$.  Similarly, for high surface
brightness galaxies, a lower $\m2l$ is required to avoid violating the
constraint $\epsm>0.63$.  Note that if $\m2l$ were not universal, then
the Tully-Fisher relation would require that $m|d\propto \m2l$.  In
the opposite case, if $\m2l$ were universal, a more stringent upper
limit on $\m2l$ would be required, in order to accommodate high
surface brightness galaxies.

\begin{figure}
\epsfxsize=.8\hsize
\computelabs
\large
\ylabel{P(\lambda)}{-0.0}{-.5}
\xlabel{\lambda}{.5}{-.95}
\normalsize
\centerline{\epsfnormal{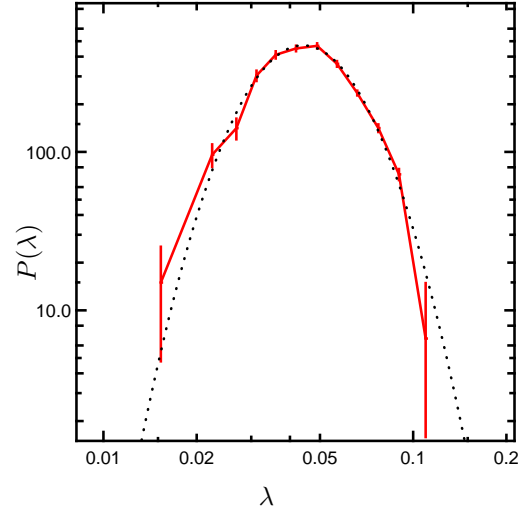}}
\vskip\baselineskip
\caption{ The distribution of $\lambda$ for the MF data (solid line with
error bars).  The best fit to a log-normal distribution ($\bar\lambda=
0.05$, $\sigma=0.36$) is shown as a dotted line.
The error bars are estimated using a standard bootstrap 
method.}
\label{distfig}
\end{figure}

\subsection{Spin parameters}\label{spinsec}

The distribution of $\lambda$ can be determined from $N$-body
simulations of hierarchical clustering, and is found to be log-normal
with mean $\bar\lambda\approx0.05$ and standard-deviation
$\sigma\approx0.5$ almost independent of cosmology (Lemson \&
Kauffmann 1998, Warren \etal{.}  1992).  If observations of disk
galaxies are a fair sample of dark matter halos, then the distribution
of $\lm$ should be closely related to the distribution of $\lambda$.
The distribution of $\lambda$ in the MF data (calculated from equation
\ref{lamchi}) is shown in Figure \ref{distfig}.  The value of
$\bar\lambda$ has been fixed at $0.05$ by choosing $\chi=0.049$.  As
discussed in Section \ref{mmwsec}, $\chi\approx0.1 H_0/H(z)$ for
isothermal halos. Using more realistic halo profiles and taking into
account disk self gravity reduces the value of $R|d$ for a given
$v|m$, which corresponds to a reduction of the value of $\chi$.
This value of $\chi$ found here is actually in good agreement with the
detailed modelling of \MMW{}.

The distribution is remarkably close to log-normal with
$\sigma=0.36\pm 0.01$. The value of $\sigma$ derived from the data is
smaller than the value, $\sigma=0.5$, given by N-body simulations of
dark halos. The reason for this discrepancy may be due to the fact
that the observational sample is biased against both low-surface
brightness galaxies, which are associated with high-spin systems
according to the disk model considered here, and early type spirals,
which are associated with low-spin systems.  The main effect of the
selection function is to rule out galaxies below some threshold in
surface brightness (corresponding to $23.5$mag/arcsec$^2$).  There is
also a distance dependent component, selecting galaxies with a range
of sizes which depends on their distance.  To check that the selection
function does not severely affect our results we constructed
Monte-Carlo samples using the model in Section \ref{mmwsec} and
applied the selection function to them.  The selection process reduced
the apparent width of the $\lambda$ distribution; the best fit to the
data came from a distribution with $\sigma_\lambda \approx 0.4$.  The
values of $\chi$ and $m|d$ (see below) are derived from $\bar\lambda$
and are little affected.  One feature of the surface brightness
threshold is that it tends to cut off the low-luminosity 
(low-$v|m$) galaxies in the Tully-Fisher relation, 
reducing the slope from the theoretical value
(3) closer to that observed ($\approx2.5$ in the MF sample).

\begin{figure}
\epsfxsize=.8\hsize
\computelabs
\large
\ylabel{m|d}{-.05}{-.5}
\xlabel{\epsl h^{-1/2}}{.5}{-.95}
\normalsize
\centerline{\epsfnormal{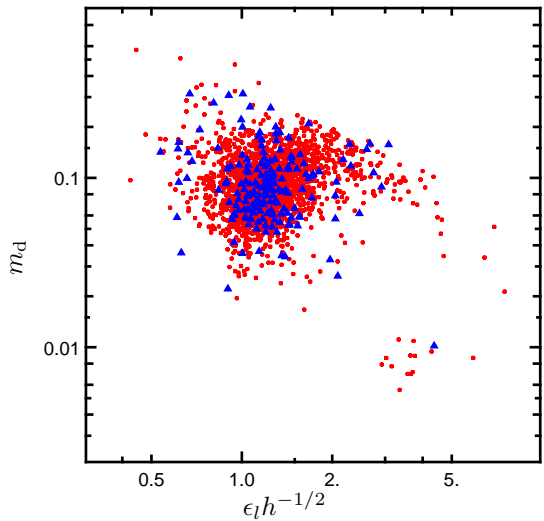}}
\vskip\baselineskip
\caption{ Shows $m|d=\chi\am \Upsilon/h$ versus $\epsl$ for the MF
data, assuming that $\m2l$ is given by the crude upper limit
$\m2l=(\epsl/0.63)^2$.  Barred galaxies are again plotted as
triangles.  }
\label{mdu}
\end{figure}

\begin{figure}
\epsfxsize=.8\hsize
\computelabs
\large
\ylabel{\bv}{-0.05}{-.5}
\xlabel{\lm}{.5}{-1.}
\normalsize
\centerline{\epsfnormal{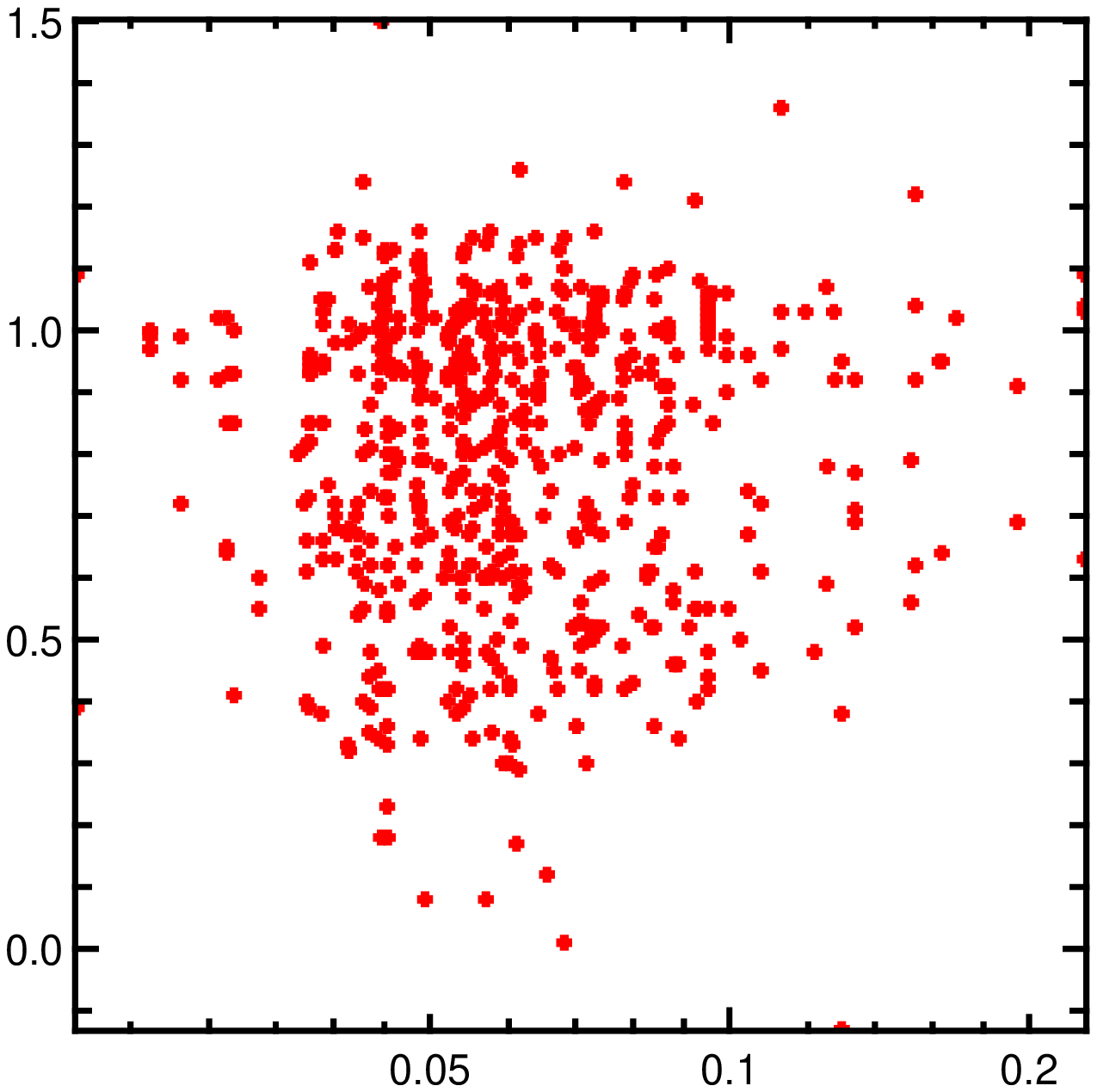}}
\vskip\baselineskip
\caption{ The MF galaxies which have $\bv$ colours in the ESO/Uppsala
catalogue.  There is a moderate scatter in $\bv$ but no trend with
$\lm$. }
\label{bvfig}
\end{figure}

\subsection{Baryon fractions}\label{mdsec}
Once $\chi$ is fixed, $a|m$ can be estimated for each galaxy using
equation (\ref{epslam}).  Restricting to those galaxies with
$\vmax>100\kms$ we find that
\begin{equation}
\langle\am\rangle = 0.47\;.
\label{amlam}
\end{equation}

{}From equations (\ref{amdef}) and (\ref{amlam}) we have the total
mass-to-light ratio in the $I$-band:
\begin{equation}
\langle M|h/L_I\rangle = {1\over\chi\langle\am\rangle} = 43h.
\end{equation}
The observed mean luminosity of the Universe 
derived from recent redshift surveys of galaxies 
gives a critical mass-to-light ratio 
$(M/L) \approx 1500h(M/L)_\odot$ to close the universe
(see e.g. Lin et al. 1996). Thus the mass associated with 
individual galactic halos gives $\Omega_{\rm gal}
=43h/1500h \sim 0.03$, which is a small fraction of the total 
mass in the universe.

Using equations (\ref{amdef}) and (\ref{amlam}) 
\begin{equation}
\langle m|d\rangle  = 0.086 \,{\m2l\over3.56h}.
\label{mdlim}
\end{equation}
The result is shown as a scatter plot in Figure \ref{mdu} using the
crude upper limit $\m2l=(\epsl/0.63)^2$.  
The gas fraction obtained for
X-ray clusters is $f_{\rm gas} \sim 0.06 h^{-3/2}$ (\eg{.} Evrard 1997).  
This fraction is usually considered to be equal to the mean value for the whole
Universe (\eg{.} White \etal{.}  1993), and therefore should be at
least as large as that in galaxies.  Equation (\ref{mdlim}) together
with the gas fraction in clusters thus
imply that
\begin{equation}
\langle \m2l\rangle  \la 2.48 \,h^{-1/2}.
\label{mdlim2}
\end{equation}
which is an new upper limit independent of that derived in Section
\ref{upssec}.  Clearly, the observational constraint on the fraction
of baryons in galactic halos has important implications for the
formation of galactic disks.
    
\section{Discussion}\label{dissec}

\subsection{Independent determinations of $\Upsilon$}\label{detsec}
The disk mass-to-light ratios we derive in Section \ref{upssec} are
formally upper limits, and based only on rotation curve data.  Thus
they are consistent with mass-to-light ratios derived by other authors
using the maximum disk hypothesis (Carignan \& Freeman 1985, Palunas
\& Williams 1998).

Limits on the mass-to-light ratio of the Galactic disk in the solar
neighbourhood can be derived from a combination of kinematic
measurements and star counts.  Kuijken \& Gilmore (1989) derive a
local surface mass density in the disk of $40\Msun/\pc^2$, and star
counts give a $V$-band luminosity density of $15\Lsun/\pc^2$ (Gould,
Bahcall \& Flynn 1996).  Dividing mass by light we obtain
$\Upsilon_V=2.67$, and thus $\m2l\approx 1.9$ (assuming $V-I=1.0$).
This number is independent of $h$, but is comparable with the median
maximum-disk value for the MF galaxies derived in Section \ref{upssec}
for $h\ga0.5$.

Mass-to-light ratios can also be derived from pure stellar population
synthesis arguments, although there is always some uncertainty arising
from the poorly known initial mass function (IMF), particularly from
the low-mass cut-off in the IMF. A few stellar population models are
available (e.g., Bertelli et al 1994; Bruzual \& Charlot 1993; Worthey
1994).  For a Salpeter IMF, the mass-to-light ratio derived by various
authors appear to agree within an accuracy of 25\% (Charlot, Worthey
\& Bressan 1996). The predicted $\m2l$ depends on the metallicity and
age of the stellar population. For a stellar population with age
between 5-12 Gyr and with a solar metallicity, $\m2l$ is between
$0.9$-$1.8$ for a constant star formation rate (\cf{} Table 3 in de
Jong 1996).  For an exponential star formation law, the mass-to-light
ratio is about 20\% higher. The predicted values are in good agreement
with the values derived in this paper. In the comparison, we have
neglected the uncertainty due to dust, since the Tully-Fisher studies
already attempt to correct for its effect.  It has been argued (de
Jong 1996) that dust reddening probably plays a minor role in the
colour gradients in disk galaxies. Nevertheless, the dust correction
remains a nuisance in these comparisons.

The mass-to-light ratio of extragalactic disks can also be measured
directly from kinematic studies such as that described by Bottema
(1993).  The measurement of $\Upsilon$ relies on the relation between
vertical velocity dispersion $\sigma_z$, surface density $\Sigma$ and
vertical scale height $z_0$.  Essentially, the larger the value of
$\Sigma$, the hotter a disk has to be at constant $z_0$.  
Using a small sample of very bright galaxies, Bottema
(1997) derives a value of $\m2l=(1.7\pm0.5)h$, which is consistent
with the upper limits we derived in Sections \ref{upssec} and
\ref{mdsec}.
It is important to extend the range of data analysed by Bottema
(1997).  This requires a good HI rotation curve for each galaxy, and
high quality spectroscopy at least along the major axis of the galaxy.
Ideally one would choose the brightest members of a large pre-defined
sample (such as that of Mathewson \& Ford 1996) and follow them up
with a high spatial resolution spectrograph.  Kinematic information in
more than one dimension is also advantageous since it removes some of
the uncertainties in the deprojection of the velocity ellipsoid.  A
number of two dimensional spectrographs are due to come on line
shortly, and these may be well suited to the problem.  The biggest
uncertainty in the determination of $\m2l$ will remain that associated
with the value of $z_0$.  Efforts should therefore be made to analyse
as many edge on galaxies as possible to try to improve on existing
determinations of the distribution of $z_0/\rd$ (\eg{.}  Barteldrees
\& Dettmar 1994).

\subsection{Uncertainties in the results}\label{sfsec}

The constant $\chi$ in equation (\ref{rhmhmmw}) might in principle
have been different owing to angular momentum loss.  The fact that the
derived value is close to the one expected in the disk formation model
implies that the gas should not have lost much angular momentum during
disk formation. This conclusion was also reached by Fall \& Estafthiou
(1980) and MMW based on the observed disk scale lengths of local
galaxies.  We confirmed this using our Monte-Carlo simulations:
decreasing the value of $\chi$ led to disks with smaller $\rd$ and
larger $\mu$ at fixed $\vmax$ compared with the MF sample ($L|d$ is
fixed by the Tully-Fisher relation).

According to MMW, when realistic halo profiles are used and disk
self-gravity included, the factor of $1/\sqrt{2}$ in equation
(\ref{rdmmw}) and the constant $\chi$ in equation (\ref{rhmhmmw})
should in principle be replaced by quantities that depend on the spin
parameter and concentration of galactic halos.  The details of these
dependences are uncertain, because they require accurate knowledge on
halo density profiles, gas settling processes and star formation
feedback.  To derive observational constraints on these details of
disk formation requires detailed decompositions of individual galaxies
into different components as well as accurate measurements of disk
rotation curves.  Such analysis is not possible with the data set used
here.

As discussed in Section \ref{upssec}, the limit on the disk
mass-to-light ratio was derived assuming that $\m2l$ is independent of
$\mu_0$ (or of $\lm$, see Figure \ref{lamufig}).  This assumption is,
as we argued, consistent with the observational data and with
independent measurements of $\m2l$ (\cf{.} Figure \ref{bvfig}).  Since
the Tully-Fisher amplitude $A$ derived from the data is quite
independent of $\mu_0$ for the majority of galaxies (\cf{.}  Figure
\ref{datfig}), any trend in $\m2l$ with $\mu_0$ has to be compensated
by a similar trend in $m|d$.  This would happen if disks with lower
surface brightness contain larger amount of gas but have a lower star
formation efficiency.  Such a trend is suggested by the observations
of McGaugh \& de Blok (1997) where gas fraction in low
surface-brightness disks is compared with that in high
surface-brightness disks.  However, for most of the galaxies in the MF
sample, disk masses are expected to be dominated by stars rather by
gas, and so the trend in the gas fraction with surface brightness
should not induce a significant trend in the disk mass-to-light ratio.

\subsection{Barred galaxies}
It is well known that isolated disks are violently unstable.  Since
disks in halos with lower spin parameters are more compact and more
self-gravitating, they are more prone to global instabilities.  One
obvious possibility is that globally unstable disks turn into barred
galaxies.  If so one might expect that barred galaxies to have
systematically smaller $\lm$ and larger $\mu_0$. No such trend is seen
in Figure \ref{lamufig} where barred galaxies (triangles) seem to be
randomly drawn from the galaxy population.  Two possibilities occur to
us.  First, if $\Upsilon$ is universal, then global instabilities in
disk galaxies must be switched off by some mechanism which is
independent of $\mu_0$.  Such a mechanism might involve central mass
concentrations (Toomre 1981) which stabilise a disk by interfering
with transmission of density waves through the centre (Sellwood \&
Moore 1998).  Second, if $\Upsilon$ depends on $\mu_0$, then it must
do so in such a way as to make all galaxies equally susceptible to bar
formation.  Bars could then be formed by interactions between galaxies
(Noguchi 1996), or between galaxies and their dark matter
halos. (N{.}B{.} the Tully-Fisher relation then requires that $m|d$
depends on $\mu_0$.)

\section*{Acknowledgments}
We are grateful to Simon White for helpful discussions.  This project
is partly supported by the ``Sonderforschungsbereich 375-95 f\"ur
Astro-Teilchenphysik'' of the Deutsche Forschungsgemeinschaft.

{}

\section*{Appendix}
Here we describe how values of $\rd$, $\mu_0$ and $L$ were derived
from the published quantities of Mathewson \& Ford (1996).  The
published data list total magnitudes, and face-on corrected isophotal
quantities: average surface brightness ($\bar \mu$) and isophotal
diameter ($R$) at $\mu$ corresponding to
$23.5 \hbox{mag/arcsec}^2$.  Assuming an
exponential disk we have a surface brightness profile
\begin{equation}
\mu(R) = \mu_0 \exp(-\alpha),~~\alpha \equiv {R \over R_d},
\label{mur}
\end{equation}
and 
\begin{equation}
L(R) = L|d \lr{1- \exp(-\alpha)(\alpha+1)}.
\label{lreq}
\end{equation}
 From equation (\ref{lreq}) the average surface
brightness inside radius $R$ is
\begin{equation}
\bar\mu = {L(R)\over\pi R^2} = {2\mu_0\over\alpha^2} 
\lr{1-\exp(-\alpha)(\alpha+1)},
\label{Iave}
\end{equation}
where we have used $L|d=2\pi\mu_0\rd^2$.  Combining equations
(\ref{lreq}) and (\ref{Iave}) we obtain
\begin{equation}
{\bar\mu\over\mu} = {\exp(\alpha) - (\alpha+1)\over \alpha^2}. 
\label{mumeq}
\end{equation}
Given $\bar\mu/\mu$ we can solve equation (\ref{mumeq})
numerically to find $\alpha$.  Then we have $\rd=R/\alpha$, and
$\mu_0=\mu\exp(\alpha)$.

\bsp
\label{lastpage}
\end{document}